\documentclass[journal]{IEEEtran}
\IEEEoverridecommandlockouts

\usepackage{cite}
\usepackage{amsmath,amssymb,amsfonts}
\usepackage{algorithmic}
\usepackage{graphicx}
\usepackage{textcomp}
\usepackage{xcolor}

\usepackage{pifont}

\usepackage{url}
\usepackage{xurl}
\usepackage[hidelinks,bookmarks=false]{hyperref}

\usepackage{tikz}
\usetikzlibrary{shapes.geometric, arrows}
\tikzstyle{startstop} = [rectangle, rounded corners, minimum width=3cm, 
    minimum height=1cm,text centered, draw=black, fill=red!20]
\tikzstyle{process} = [rectangle, minimum width=3cm, minimum height=1cm, 
    text centered, draw=black, fill=blue!20]
\tikzstyle{decision} = [diamond, minimum width=3cm, minimum height=1cm, 
    text centered, draw=black, fill=green!20]
\tikzstyle{arrow} = [thick,->,>=stealth]

\def\BibTeX{{\rm B\kern-.05em{\sc i\kern-.025em b}\kern-.08em
    T\kern-.1667em\lower.7ex\hbox{E}\kern-.125emX}}

\begin{document}


\title{EduSOC: A Lightweight Simulated Security Operations Center with AI Incident Responders}



\author{Martin Higgins, Shawn Thompson, Cherry Mangla, and Bobby Swar%
\thanks{The authors are with the Faculty of Management,
Concordia University of Edmonton, Edmonton, AB, Canada
(e-mail: martin.higgins@concordia.ab.ca;
shawn.thompson@concordia.ab.ca;
cherry.mangla@concordia.ab.ca;
bobby.swar@concordia.ab.ca).}%
}




\maketitle


\begin{abstract}
This paper presents \emph{EduSOC}, a lightweight web-based Security Operations Center (SOC) simulation platform for cybersecurity education. Conventional cyber ranges can be complex and resource-intensive to deploy, particularly for introductory exercises focused on investigative decision-making rather than infrastructure configuration. EduSOC provides an alternative by combining continuous synthetic event generation, searchable supporting evidence, and role-based artificial intelligence (AI) responders that assist investigations without revealing underlying incident classifications. Students remain responsible for interpreting evidence, consulting simulated operational teams, and justifying their escalation decisions. 
A preliminary classroom evaluation was conducted across three university cohorts, comprising 50
matched pre- and post-simulation survey participants. The findings indicated high student engagement and improvements in self-reported SOC self-efficacy, while short-term knowledge gains varied across cohorts. These findings provide empirical insight into the use of lightweight, AI-supported SOC simulations in university cybersecurity education.

\end{abstract}


\section{Introduction}
Security Operations Centers (SOCs) play an important role in defending organisations against cyber threats. Analysts typically work with Security Information and Event Management (SIEM) platforms, which aggregate security events and generate alerts for investigation. An alert does not necessarily indicate a genuine attack, and analysts must often work with incomplete or conflicting information before determining whether an incident requires further action. This process can involve consultation with other operational teams and requires judgement beyond simply responding to the severity assigned to an alert \cite{b0}.

Introducing students to this environment presents a practical challenge. Cyber ranges and enterprise security platforms can provide realistic training experiences, but their deployment often requires considerable infrastructure and technical expertise \cite{b12}. For students encountering SOC operations for the first time, the complexity of configuring and maintaining such environments may also distract from the investigative process itself. There is therefore a case for simpler teaching environments that allow students to experience the decisions faced by SOC analysts without reproducing the entire technical infrastructure.

We developed \emph{EduSOC} to address this problem. The platform generates a continuous stream of simulated security alerts and provides students with access to supporting evidence through a browser-based interface. Students investigate these alerts and determine how they should be handled, including whether further investigation or escalation is warranted. Where additional information is needed, they can communicate with AI responders representing different operational teams. These responders are designed to provide information relevant to the investigation without disclosing whether the underlying event is a genuine incident or a false positive. The student consequently remains responsible for reaching and justifying a conclusion. In doing so, EduSOC allows students to experience the investigative demands of SOC operations without first having to develop the technical expertise needed to configure and operate enterprise security infrastructure.

\subsection{Novel Contributions}

The principal contributions are:

\begin{itemize}
    \item A lightweight, self-hosted SOC simulation platform that allows multiple student teams to investigate a shared stream of security events without requiring enterprise security infrastructure.

    \item An evidence-based investigation environment in which synthetic alerts are accompanied by reproducible supporting records, allowing students to practise incident classification, prioritisation, and escalation.

    \item A role-based AI communication approach that separates programmatically generated operational findings from constrained language-model responses, encouraging students to investigate evidence rather than receive definitive answers.

    \item A classroom evaluation involving 50 matched participants across three cohorts, examining short-term changes in SOC knowledge and self-efficacy alongside student engagement, usability, and qualitative feedback.
\end{itemize}

\section{Background and Related Work}

\subsection{Cyber Ranges and SOC Training}

Cyber ranges have become an established approach to practical cybersecurity education, allowing students to investigate security incidents in controlled environments without exposing operational systems to risk \cite{yamin2020,katsantonis2023}. Research has examined their use across different educational settings, including the challenges of providing authentic training experiences for learners with varying levels of technical expertise \cite{glas2023,lazarov2026}.

The relationship between technical realism and educational value remains an important consideration. More complex environments can provide authentic operational experiences, but the demands placed on learners must be appropriate to the intended learning objectives \cite{glas2024,katsantonis2023}. Deployment is another consideration, particularly where teaching staff have limited infrastructure or technical support. Even open-source cyber ranges vary considerably in their configuration requirements and documentation \cite{lundqvist2025}.

\subsection{Simulation-Based Learning and AI-Supported Instruction}

Research on simulation-based cybersecurity education has explored ways of providing practical exercises that are accessible to students with different levels of experience. Vekaria \emph{et al.} \cite{vekaria2021} demonstrated a cyber-range approach to attack-and-defence learning, while Vykopal \emph{et al.} \cite{vykopal2023} examined adaptive training environments. Other work has explored purpose-built virtual laboratories and scenario-driven activities as alternatives to conventional technical exercises \cite{yucel2024,b1}.

Large language models introduce another possibility for these environments. Rather than relying entirely on predefined interactions, students can engage with simulated participants who respond to questions within the context of an exercise. Research on educational AI agents has examined the potential for such interactions to support learning \cite{b16}. In SOC training, this approach could allow students to obtain information from simulated operational teams while retaining responsibility for interpreting the evidence themselves.

Several studies have also evaluated purpose-built cybersecurity simulations in educational settings. Jaffray \emph{et al.} \cite{b11} developed SherLOCKED, a detective-themed serious game used to reinforce introductory cybersecurity concepts, while Mason \emph{et al.} \cite{b7} evaluated a gamified cybersecurity training environment with students from non-IT backgrounds. More technically focused environments have also incorporated adaptive scenarios and learning analytics. These studies illustrate the different levels of technical complexity at which practical cybersecurity education can be delivered.



\subsection{Positioning EduSOC Among Existing Platforms}
Several existing platforms provide environments for practical SOC training. RangeForce \cite{rangeforce} and Cloud Range \cite{cloudrange} offer cyber-range exercises involving operational security tools, while TryHackMe SOC Simulator \cite{tryhackmesoc} provides browser-based activities involving alert investigation and case reporting. OpenSOC \cite{opensoc} takes a different approach, emphasising enterprise network simulation. These platforms demonstrate the range of approaches available for SOC training, from guided investigations to more comprehensive technical environments.

EduSOC takes a narrower approach, concentrating on the investigative decisions faced by SOC analysts rather than reproducing enterprise security infrastructure. A distinguishing feature is the separation of deterministic incident evidence from AI-mediated operational communication. Students investigate simulated alerts and consult role-based AI responders that provide contextual information without revealing definitive incident classifications. The event stream can be controlled by the instructor, with separate student groups investigating the same scenarios independently. This approach allows SOC exercises to be delivered within a conventional classroom setting, with relatively modest infrastructure requirements while retaining the need for students to interpret evidence and justify their decisions.

\section{Educational Design Rationale}

\subsection{Intended Learning Outcomes}
EduSOC was designed to introduce students to the investigative responsibilities of SOC analysts within a single classroom exercise. Rather than focusing on the configuration of security infrastructure, the simulation places students in situations where they must interpret potentially suspicious activity and determine an appropriate response. Students are expected to distinguish genuine incidents from false positives, assess the available evidence, and justify their decisions about prioritisation and escalation. An important objective is to develop an appreciation of the uncertainty involved in SOC investigations, where the severity assigned to an alert may not accurately reflect the underlying threat.

The exercise also introduces the communication involved in incident investigation. Students can consult simulated operational teams through role-based AI responders when additional information is required, but remain responsible for reaching their own conclusions. By the end of the exercise, students should be able to document an investigation, explain the evidence supporting their assessment, and identify when an incident should be escalated.

\subsection{Classroom Integration and Workflow}

EduSOC is designed for an instructor-led exercise lasting approximately 90 minutes. Following a short briefing on SOC operations, students are assigned to teams and access separate simulation rooms through a web browser. Each team receives the same stream of synthetic security alerts while maintaining independent investigation records and communications.

Students monitor a SIEM-style dashboard, select alerts for investigation, examine supporting evidence, and consult AI-simulated operational teams where additional context is required. They then submit their classification, confidence level, supporting rationale, and recommended action. This workflow encourages evidence-based triage and escalation decisions rather than reliance on alert severity alone.

The instructor oversees the exercise through a separate interface, with controls to adjust event generation, introduce incidents, and monitor student activity. The session concludes with a debrief in which students discuss their decisions and reflect on the challenges of investigating uncertain security events.

\section{System Overview}

\begin{figure*}[t]
    \centering
    \includegraphics[width=0.95\textwidth]{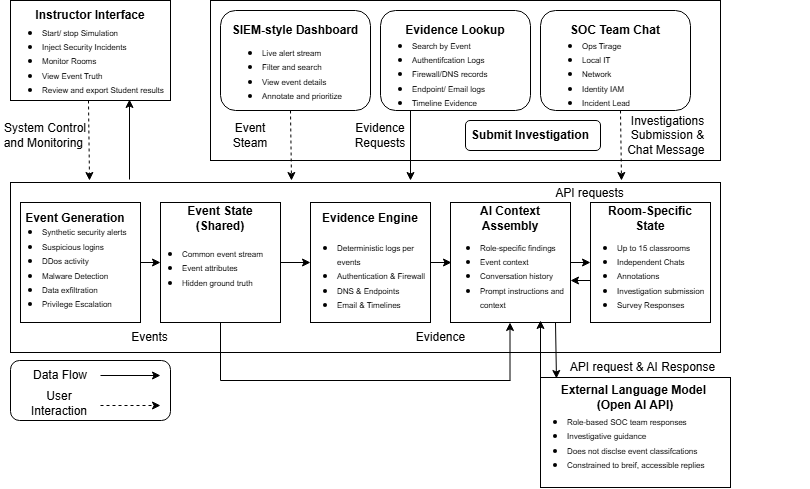}
    \caption{High-level architecture of EduSOC, illustrating the instructor and student interfaces, synthetic event generation, shared event state, deterministic evidence retrieval, room-specific investigation management, and role-based AI communication through an external language-model service.}
    \label{fig:architecture}
\end{figure*}

\subsection{System Architecture and Implementation}

EduSOC is implemented using a Python Flask-SocketIO backend and a lightweight browser-based front end. The application integrates synthetic event generation, searchable supporting evidence, real-time communication, and AI-assisted investigation within a single locally hosted environment. Its architecture is designed to operate on modest computing hardware without requiring enterprise security monitoring infrastructure.

A background process continuously generates synthetic security alerts representing suspicious logins, denial-of-service activity, malware detection, data exfiltration, and privilege escalation. Each event contains attributes including timestamp, source IP address, region, device, and severity, alongside a hidden ground-truth classification identifying genuine incidents and false positives. Supporting authentication, firewall, DNS, endpoint, email, and timeline records are generated deterministically from the event state, providing reproducible investigative evidence without requiring live network traffic.

The platform supports up to 15 classroom rooms receiving a common event stream while maintaining independent chat histories, annotations, and investigation records. Flask-SocketIO enables real-time communication within each room, allowing multiple student teams to investigate the same events independently without deploying separate application instances.

The core simulation and evidence-generation components operate locally, while AI-assisted communication requires access to the external language-model service.

\subsection{AI Incident Responders}
A central design decision in EduSOC was to use large language models as simulated operational colleagues rather than as tutors or automated incident classifiers. Allowing a general-purpose chatbot to interpret an alert directly could undermine the exercise by providing students with the answer. Conversely, relying entirely on scripted dialogue would restrict the questions students could ask and require extensive preparation for each scenario. We therefore adopted a role-based prompting approach in which AI responders provide operational information while leaving the investigation to the student.

The system uses a common prompt defining five operational roles: Ops Triage, Local IT, Network, Identity and Access Management (IAM), and Incident Lead. Each role has defined responsibilities and limitations on the information it can provide. For example, the Network responder can discuss firewall activity and unusual outbound connections but cannot confirm account ownership, while IAM can examine authentication activity and privilege changes but cannot investigate malware on an endpoint. Students must consequently identify which team is likely to possess the information needed for their investigation.

The prompt engineering in EduSOC was designed around an educational constraint: AI responders should support the investigative process without performing the investigation for the student. Rather than deploying a general-purpose chatbot, the system assigns operational responsibilities to five simulated teams and supplies each with role-specific findings generated from the underlying event state. Students must identify the appropriate team, request relevant information, and independently interpret the evidence. The language model provides a conversational interface to the scenario, while the simulator retains control over the underlying incident and its supporting evidence. The prompts also constrain how information is communicated. Responders are instructed to direct students towards relevant evidence sources rather than provide a definitive interpretation. One instruction explicitly requires the model to ``direct the student to the appropriate evidence category rather than quoting the decisive log line.'' In a suspicious-login scenario, for example, IAM may report repeated authentication failures followed by an unusual successful login and suggest examining the authentication records. The student must then determine whether the evidence warrants escalation.

To maintain consistency, EduSOC generates scenario-specific findings programmatically according to the event type, underlying classification, and operational role. These findings are supplied to the language model alongside the selected event and recent conversation history. The prompt prohibits the responder from disclosing the true classification or confirming whether the student's conclusion is correct. Responses are also kept brief and generally conclude with a suggested investigative action.

\subsection{Student and Instructor Interfaces}

\begin{figure*}[t]
    \centering
    \includegraphics[width=1.0\textwidth]{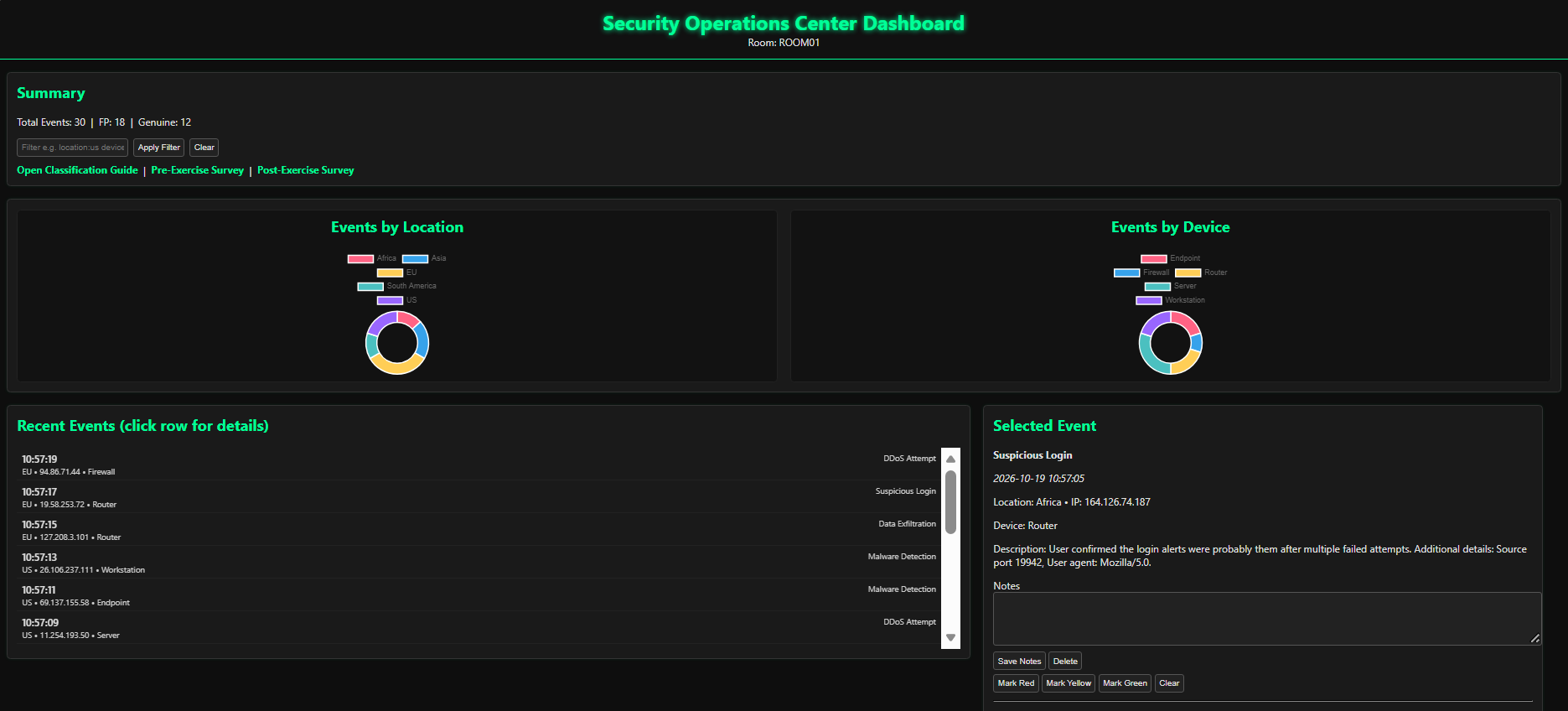}
    \caption{Student dashboard overview. The interface presents the simulated event stream, summary visualizations, filtering controls, event details, investigation tools, and access to role-based SOC team communication.}
    \label{fig:student-top}
\end{figure*}

The Student Dashboard provides a SIEM-style view of incoming security events, including severity information, filtering controls, summary visualizations, and event details, as shown in Fig.~\ref{fig:student-top}. Students can annotate alerts, examine supporting evidence, and submit investigations containing their classification, confidence level, rationale, and recommended action.

\begin{figure}[t]
    \centering
    \includegraphics[width=0.5\textwidth]{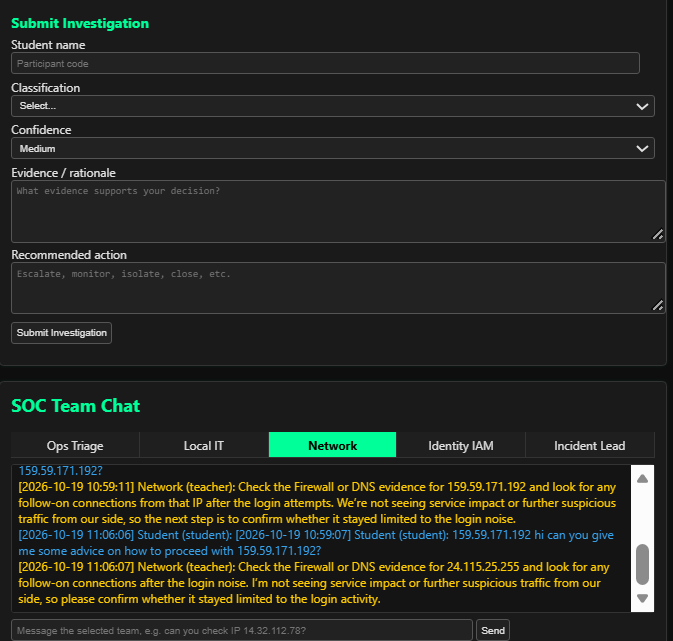}
    \caption{EduSOC student dashboard showing the structured investigation submission panel for recording classifications, confidence, supporting evidence, and recommended actions, and an example of role-based communication with simulated SOC operational teams.}
    \label{fig:airesponder}
\end{figure}

Operational communication is supported through five role-based AI responders: Ops Triage, Local IT, Network, Identity and Access Management (IAM), and Incident Lead. Fig.~\ref{fig:airesponder} illustrates the investigation submission interface alongside an example of a simulated operational conversation.

\begin{figure}[t]
    \centering
    \includegraphics[width=0.5\textwidth]{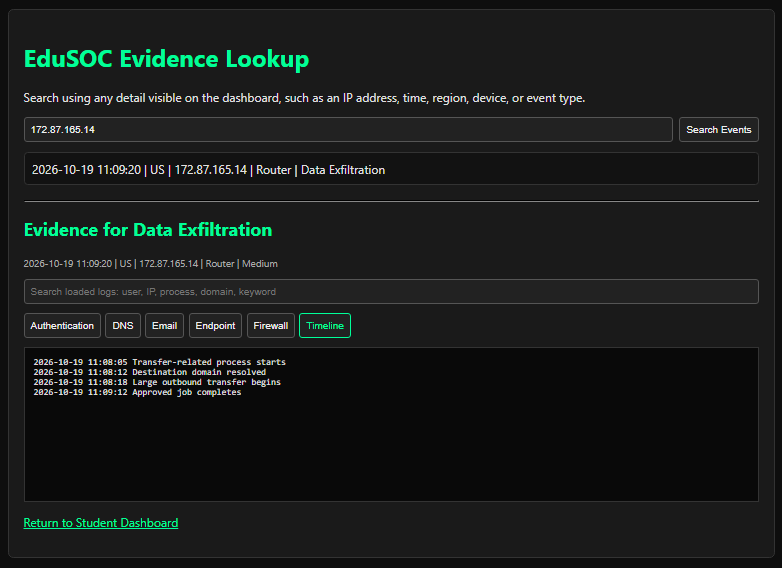}
    \caption{EduSOC Evidence Lookup dashboard showing event search functionality and simulated security logs across authentication, DNS, email, endpoint, firewall, and timeline sources. The example illustrates timeline evidence associated with a potential data exfiltration alert.}
    \label{fig:evidence}
\end{figure}

The Evidence Lookup provides searchable authentication, firewall, DNS, endpoint, email, and timeline records associated with simulated events. Fig.~\ref{fig:evidence} illustrates how students can retrieve additional evidence when investigating a potential security incident.

A separate Instructor Dashboard allows the instructor to monitor classroom activity, control event generation, inject additional incidents, review hidden event classifications, and export investigation and survey records.

\subsection{System Design Challenges and Mitigations}

\subsubsection{University Network Deployment} A central challenge in developing EduSOC was balancing ease of deployment with the functionality required for a realistic classroom exercise. University network restrictions can complicate hosting applications that require continuous communication between students and a central server. EduSOC addresses this through a lightweight architecture that can operate on a local network without requiring enterprise infrastructure. During classroom trials, the platform was also deployed successfully using Cloudflare Tunnel, enabling remote access to a locally hosted instance without requiring changes to institutional firewall configurations.

\subsubsection{Supporting Big Cohorts} Supporting multiple student groups simultaneously presented a further design consideration. Rather than deploying separate instances for each group, EduSOC uses a shared event stream with room-specific investigation records and AI conversations. This allows teams to investigate the same incidents independently while limiting infrastructure requirements. Instructors can also adjust event generation to manage the volume of incoming alerts and accommodate differences in student experience.

\subsubsection{Meaningful Investigation} Another challenge was providing sufficient investigative detail without reproducing an entire enterprise network. EduSOC addresses this through programmatically generated security events and supporting evidence, allowing students to examine consistent authentication, network, and endpoint records without requiring live traffic capture or complex security tooling. Although this approach sacrifices some technical realism, it preserves the investigative activities central to the exercise.

\subsubsection{Balancing AI Assistance and Student Autonomy}

Integrating AI responders presented a particular design challenge. Allowing students to interact directly with a general-purpose language model could undermine the learning objectives if it simply identified genuine incidents or provided the correct investigative conclusions. Furthermore, exposing the underlying API directly to students would introduce risks associated with credential security, uncontrolled usage, and potentially excessive API costs. EduSOC therefore routes AI interactions through its backend, allowing the application to manage requests and supply relevant scenario information without exposing API credentials to students.

Considerable attention was also given to prompt engineering and behavioural guardrails. Responders are assigned specific operational roles, with instructions limiting the information they can provide and preventing them from explicitly revealing incident classifications or confirming student answers. Rather than resolving investigations, they are directed to suggest relevant evidence sources and encourage students to explain their own conclusions. This approach allows AI to reduce the burden on instructors who would otherwise need to represent multiple operational teams, while preserving the student's responsibility for investigative decision-making. Nevertheless, these prompt-based restrictions cannot guarantee consistent behaviour, making continued testing and refinement necessary.

\section{Study Design}
EduSOC was evaluated using a pre- and post-simulation survey design across three university classroom cohorts. \textbf{BUS 250 }comprised primarily undergraduate management students enrolled in Management Information Systems, a course covering the role of information technology in business operations and organisational decision-making. \textbf{ISSM 551 }comprised postgraduate information security and assurance students studying disaster recovery and business continuity, while \textbf{CSM 303A} consisted of undergraduate information technology and management students enrolled in a security operations management course. The cohorts therefore represented different academic backgrounds and levels of formal exposure to cybersecurity. Participants completed an instructor-led SOC simulation exercise lasting approximately 90 minutes, working in small groups to investigate simulated security alerts and communicate with operational teams.

The survey assessed SOC-related knowledge using ten scenario-based multiple-choice questions covering alert interpretation, investigative decision-making, prioritisation, and escalation. The same questions were administered before and after the exercise, with knowledge scores calculated as the number of correct responses out of ten. Students also completed five self-efficacy items assessing their perceived ability to investigate, prioritise, and justify SOC decisions. Responses were recorded on a five-point Likert scale and averaged to produce an overall self-efficacy score.

Following the exercise, participants completed additional four-item measures concerning engagement, usability, and workload-related perceptions, also using five-point response scales. Three open-ended questions invited students to reflect on what helped their investigations, the difficulties they encountered, and whether their understanding of SOC operations had changed.

Surveys were administered primarily through EduSOC's online interface, with paper questionnaires retained as a backup. Paper responses were used to supplement incomplete online records in the undergraduate management cohort. Pre- and post-simulation responses were matched using participant-provided identifiers, with cohort and
survey-source information used to resolve discrepancies where necessary. The final matched sample comprised 50 participants with usable pre- and post-simulation knowledge and self-efficacy responses.

Changes in knowledge and self-efficacy were assessed using two-tailed paired-sample $t$-tests comparing matched pre-and post-simulation scores. Results were examined both across the full sample and descriptively by cohort. Effect sizes were calculated using Cohen's $d_z$. Statistical significance was assessed at the 0.05 level.

The study employed a within-participant pre- and post-simulation design to assess changes in SOC knowledge and self-efficacy. The pre-simulation surveys established a baseline of participants' existing SOC knowledge and confidence prior to using EduSOC, against which post-simulation outcomes could be compared. By matching individual responses, the evaluation measured changes in knowledge and self-efficacy across students with different academic backgrounds and levels of prior cybersecurity experience.

\paragraph{Research Ethics}
The classroom evaluation was conducted under research ethics approval Pro00159393, obtained through the University of Alberta and Concordia University of Edmonton research shared ethics process (ARISE). Survey participation was voluntary, and informed consent was indicated through survey completion and submission. Pre- and post-simulation responses were matched using participant-provided identifiers. These identifiers were used to facilitate survey matching, with the resulting data analysed in de-identified form.

\section{Results}

\subsection{Participants and Survey Matching}
The analysis included 50 matched pre- and post-simulation survey responses across three cohorts: 11 undergraduate management students (BUS 250), 19 postgraduate information security students (ISSM 551), and 20 undergraduate information technology/management students (CSM 303A). Matched knowledge and self-efficacy scores were available for all 50 participants, alongside post-simulation engagement, usability, and workload-related ratings.

\subsection{Knowledge Acquisition}

Across the 50 matched participants, mean knowledge scores increased from 6.50 to 7.38 out of 10. A paired-sample $t$-test indicated a statistically significant improvement, $t(49)=3.12$, $p=0.0031$, with a moderate within-participant effect size ($d_z=0.44$). The mean improvement was 0.88 points (95\% CI [0.31, 1.45]). However, exploratory cohort-level analyses indicated that the increase was concentrated among postgraduate security students.

Postgraduate security/assurance students demonstrated the largest increase, from 6.00 to 7.95 (+1.95 points), with a statistically significant improvement ($t(18)=4.22$, $p<0.001$, $d_z=0.97$). In contrast, management undergraduates improved from 7.18 to 7.27 (+0.09), while IT/management undergraduates increased from 6.60 to 6.90 (+0.30), neither of which represented a statistically significant change ($p=0.858$ and $p=0.460$, respectively). Thus, the overall improvement was driven primarily by the postgraduate cohort, with relatively limited changes observed among undergraduates.

These results indicate short-term improvements in performance on the knowledge assessment, although they do not independently establish improved operational competence.

\begin{table}[t]
\centering
\caption{Matched Pre- and Post-Simulation Knowledge Scores}
\label{tab:knowledge}
\footnotesize
\setlength{\tabcolsep}{3pt}
\begin{tabular}{lrrrr}
\hline
Cohort & $n$ & Pre ($M\pm SD$) & Post ($M\pm SD$) &
$\Delta$ ($M\pm SD$) \\
\hline
Mgmt. UG     & 11 & $7.18\pm2.52$ & $7.27\pm3.00$ & $+0.09\pm1.64$ \\
Security PG  & 19 & $6.00\pm2.13$ & $7.95\pm1.96$ & $+1.95\pm2.01$ \\
IT/Mgmt. UG  & 20 & $6.60\pm2.41$ & $6.90\pm2.59$ & $+0.30\pm1.78$ \\
\hline
Overall      & 50 & $6.50\pm2.33$ & $7.38\pm2.47$ & $+0.88\pm2.00$ \\
\hline
\end{tabular}
\end{table}

\subsection{SOC Self-Efficacy}

Self-efficacy increased across all three cohorts, with the overall mean rising from 3.17 to 3.56 on a five-point scale (+0.38; Table~\ref{tab:selfefficacy}). A paired-sample $t$-test indicated a statistically significant improvement, $t(49)=4.05$, $p<0.001$, with a moderate within-participant effect size ($d_z=0.57$). The mean improvement was 0.38 points (95\% CI [0.19, 0.57]).

Management undergraduates showed the greatest increase (+0.62, $p=0.0358$), followed by IT/management undergraduates (+0.37, $p=0.0351$) and postgraduate security/assurance students (+0.26, $p=0.0244$). All three cohorts demonstrated statistically significant improvements in self-efficacy at the nominal 5 percent significance level. Unlike the knowledge assessment, where significant improvement was observed primarily among postgraduate students, self-reported confidence increased significantly across all three cohorts.

These findings suggest an improvement in participants' perceived ability to undertake SOC-related activities. However, self-efficacy represents reported confidence rather than independently assessed operational performance.

\begin{table}[t]
\centering
\caption{Matched Pre- and Post-Simulation SOC Self-Efficacy}
\label{tab:selfefficacy}
\footnotesize
\setlength{\tabcolsep}{3pt}
\begin{tabular}{lrrrr}
\hline
Cohort & $n$ & Pre ($M\pm SD$) & Post ($M\pm SD$) &
$\Delta$ ($M\pm SD$) \\
\hline
Mgmt. UG     & 11 & $3.02\pm0.98$ & $3.64\pm0.56$ & $+0.62\pm0.85$ \\
Security PG  & 19 & $3.29\pm1.02$ & $3.56\pm1.11$ & $+0.26\pm0.47$ \\
IT/Mgmt. UG  & 20 & $3.14\pm0.55$ & $3.51\pm0.72$ & $+0.37\pm0.73$ \\
\hline
Overall      & 50 & $3.17\pm0.84$ & $3.56\pm0.85$ & $+0.38\pm0.67$ \\
\hline
\end{tabular}
\end{table}

\subsection{Student Experience}

Post-simulation engagement was high overall ($M=4.53$, $SD=0.72$), while usability was rated positively ($M=3.95$, $SD=0.83$), as shown in Table~\ref{tab:experience}. Engagement scores were relatively consistent across cohorts, ranging from 4.38 to 4.63. Workload-related ratings averaged 3.80 ($SD=0.76$) across the matched participants.

Informal feedback during the classroom sessions was also generally positive, with several students expressing interest in accessing the platform outside scheduled teaching activities. Although these observations were not systematically measured, they provide additional context for the positive engagement ratings.

\begin{table}[t]
\centering
\caption{Post-Simulation Student Experience Among Matched Participants}
\label{tab:experience}
\footnotesize
\setlength{\tabcolsep}{4pt}
\begin{tabular}{lrrr}
\hline
Cohort & Engagement & Usability & Workload \\
\hline
Mgmt. UG    & $4.61\pm0.63$ & $4.30\pm0.73$ & $3.59\pm0.78$ \\
Security PG & $4.38\pm0.92$ & $3.70\pm1.00$ & $3.86\pm0.94$ \\
IT/Mgmt. UG & $4.63\pm0.53$ & $4.00\pm0.65$ & $3.86\pm0.55$ \\
\hline
Overall     & $4.53\pm0.72$ & $3.95\pm0.83$ & $3.80\pm0.76$ \\
\hline
\multicolumn{4}{l}{\scriptsize Values are $M\pm SD$; $n=50$ (11, 19, and 20 by cohort).} \\
\end{tabular}
\end{table}

\subsection{Qualitative Feedback}
Open-ended feedback was available from 50 participants and examined using preliminary descriptive coding. Common themes included evidence interpretation (30), collaboration (24), interface and learning support (20), prioritization and escalation (17), and time pressure and alert volume (15). Students frequently described SOC operations as more complex and collaborative than anticipated, particularly when distinguishing genuine incidents from false positives and making decisions under time pressure.

Feedback on the AI incident responders was mixed. Some students valued their investigative guidance, with one noting that ``the chatbots were useful, they guided me on what to search.'' Others reported difficulty selecting the appropriate responder or receiving vague and inconsistent answers. These observations suggest that AI responders can support investigative activities but require further refinement to provide reliable, contextually appropriate guidance without replacing student judgement.

Overall, the qualitative feedback supports the positive engagement ratings while highlighting the challenges of simulating realistic SOC decision-making.

\section{Discussion \& Limitations}
The classroom evaluation produced encouraging results, although learning outcomes differed across cohorts. Self-efficacy increased across all three groups, while knowledge gains were concentrated among postgraduate security and assurance students. EduSOC requires students to interpret evidence, consult simulated operational teams, and make independent escalation decisions. Students with more specialised cybersecurity knowledge may have been better positioned to benefit from this investigative approach, while less experienced students may require additional guidance to translate their increased confidence into measurable knowledge gains. Differences in prior knowledge, course content, and familiarity with the assessment may also have contributed to these outcomes.

The qualitative feedback provides further insight into how students experienced the simulation. Students frequently identified evidence interpretation, collaboration, and prioritisation as important aspects of the exercise, suggesting that the platform exposed them to some of the practical demands of SOC operations. The AI responders were useful to some participants in identifying where to look for additional evidence, although others experienced difficulty obtaining relevant responses. This highlights an important design consideration: AI-supported operational communication can provide investigative guidance without requiring constant instructor intervention, but its usefulness depends on whether the responses are sufficiently reliable and appropriate to the student's level of understanding.

From a technical perspective, EduSOC relies on synthetic security events and supporting evidence rather than live network traffic, limiting the depth and realism of investigations compared with operational SOC environments. Although the core simulation can operate locally, the AI responders require integration with an external language-model service. Instructors must configure an OpenAI API key, maintain internet connectivity, and manage API usage costs, introducing an additional technical and financial barrier to deployment. The use of an external service also creates a dependency on its availability and response performance, potentially affecting classroom exercises. Finally, although role-based prompts are designed to prevent AI responders from disclosing incident classifications or providing definitive answers, these behavioural restrictions cannot guarantee consistent responses. Further testing is required to assess their reliability across different scenarios and student interactions.

The classroom evaluation was intended to provide an initial assessment of EduSOC in practical teaching settings rather than a controlled measurement of educational effectiveness. Its deployment across three university courses demonstrated the feasibility of incorporating the platform into classroom activities, while the survey findings provided evidence of positive student engagement and changes in self-efficacy and short-term knowledge performance. The use of matched pre- and post-simulation assessments allowed changes in knowledge and self-efficacy to be examined relative to each student's initial baseline. Although the study did not include a separate comparison group or delayed follow-up, this approach provided an initial measure of short-term learning outcomes across three distinct classroom cohorts. Future evaluations could build upon these findings through longer-term assessments and comparisons with alternative teaching approaches.

\section{Future Work}

Future development will extend EduSOC beyond initial alert investigation and triage to incorporate a broader range of SOC operations. This will include simulated operational control functions, allowing students to coordinate containment and response actions, alongside forensic feedback from affected systems to assess the consequences of their decisions. Other planned improvements include refined role-specific prompts, stronger behavioural constraints, and systematic evaluation of AI responses to ensure useful guidance without revealing incident classifications or replacing student judgement. These additions would enable more complex, multi-stage investigations in which students follow incidents from initial detection through escalation, response, and recovery. Future classroom evaluations will examine how these expanded capabilities influence investigative performance, collaboration, and knowledge retention.

\section{Conclusion}

This paper presented EduSOC, a lightweight Security Operations Center simulation platform designed to introduce students to cybersecurity incident investigation without requiring enterprise security infrastructure. By combining synthetic security events, searchable supporting evidence, and role-based AI incident responders, the platform allows students to practise alert triage, investigative communication, and escalation decisions within an instructor-led classroom exercise.

An initial evaluation across three university cohorts involving 50 matched participants found high levels of student engagement and improvements in self-reported SOC self-efficacy. Knowledge gains were less consistent, with the largest improvements observed among postgraduate information security students. These findings suggest that EduSOC provides an engaging environment for introductory SOC training, although further evaluation is required to establish its effects on learning and operational competence.


\section*{Demonstration Availability}
A beta demonstration and user guide are available at \url{https://edusoc.prairie-dog.ca}.
Access credentials are provided separately through the manuscript submission system.

\section*{Acknowledgments}
OpenAI's API was integrated into the EduSOC simulation platform to provide the AI incident responder functionality. OpenAI tools, were also used to assist with software development, code correction, debugging, and error detection, as well as language editing and manuscript organisation. These tools were not used to independently generate research findings, results or draw conclusions.  The authors would like to thank Concordia University of Edmonton for supporting this work.

\end{document}